# Weak value amplification using spectral interference of Fano resonance

Shyamal Guchhait[+], Athira B S[+], Niladri Modak[+], Jeeban K Nayak[+],
Anwesha Panda, Mandira Pal and Nirmalya Ghosh

[1]*Department of Physical Sciences,*

*Indian Institute of Science Education and Research (IISER) Kolkata.*

*Mohanpur 741246, India*

[2]*Center of Excellence in Space Sciences India,*

*Indian Institute of Science Education and Research (IISER) Kolkata.*

*Mohanpur 741246, India*

*+: these authors have contributed equally to this work*

**Abstract**

The extraordinary concept of weak value amplification and a number of other seemingly counter-intuitive wave phenomena are manifestations of fine interference effects. Taking example of one such intriguing wave interference phenomenon, namely, optical Fano resonance, we show that weak value amplification *naturally* arises in such scenario due to *near* destructive spectral interference between a continuum mode and a narrow resonance mode which mimics the corresponding *near* destructive interference between the eigenstates of the measuring observable as a consequence of *nearly* mutually orthogonal pre and post selections of the system states. In order to elucidate this, we first experimentally demonstrate a weak measurement concept that uses near destructive interference of two paths of an interferometer having slightly rotated linear polarization states of light.  By generating *nearly* destructive interference between the two paths with small amplitude and phase offsets of the waves, we observe both real and imaginary (respectively) weak value amplification of the small polarization rotation leading to large rotation of the polarization vector orientation and dramatic changes in the circular polarization descriptor $4^{th}$ Stokes vector element of light**,** respectively. We go on to demonstrate that the giant enhancement of small Faraday rotation in the vicinity of the Fano spectral dip in waveguided magneto-plasmonic crystal is a manifestation of such *natural* interferometric weak value amplification.

The weak measurement concept, introduced by Aharonov, Albert, and Vaidman [1-5], involves three steps, quantum state preparation (pre-selection), a weak coupling between the pointer (device) and the measuring observable, and post-selection on a final state which is *nearly* orthogonal to the initial state. The outcome, the so-called weak value may lie far outside the eigenvalue spectrum of an observable and can also assume complex values. These strange characteristics have allowed a wide range of applicability of weak values in both classical and quantum contexts [6-15]. The weak value amplification (WVA) has turned out to be a useful tool for addressing foundational questions in quantum mechanics [12, 13] and for resolving quantum paradoxes [14, 15]. WVA is also finding widespread metrological applications [6-11], to quantify small physical parameters, e.g., for precision measurements of angular rotation [6], phase shift [8], temporal shift [9], frequency shift [10], detection of ultrasensitive beam deflections [11], and so forth.

Even though WVA is a quantum mechanical concept, it can be understood using the wave interference phenomena and can therefore be realized in classical optical setting also [2,3]. Indeed, most of the experiments on weak measurement till date have been performed in the classical optics domain [3,6-11]. In most of these optical weak measurements, Gaussian spatial modes of laser beams or Gaussian temporal pulse are employed as external pointer and polarization state of light is conveniently used as a pre-post selection mechanism, with tiny polarization dependent optical effects providing the desired weak coupling between the pointer and the measuring observable [3,6,7,9,10]. The enigmatic concept of WVA can also be interpreted as nearly destructive interference between the eigenstates of the measuring observable as a consequence of nearly mutually orthogonal pre and post selections of the system states. There are a number of interesting, intricate and seemingly counter-intuitive phenomena in both quantum matter waves and classical electromagnetic waves that originate due to such fine interference effects, e.g., Fano resonance, Electromagnetically induced transparency (EIT) and absorption (EIA), super and sub-luminal propagation of wave packets, coherent perfect absorption and super scattering etc. [16-19]. The important question that we address here is that is it possible to integrate the concept of weak measurement with the aforementioned intriguing wave phenomena using the common platform of interference so that the weak value amplification of any weak interaction parameter may evolve naturally in such scenario. Taking an example of optical Fano resonance [20-23] in waveguided magneto-plasmonic crystal [24-26], we show that spectral domain Fano interference between a continuum mode and a narrow resonance mode having slightly different polarization response can indeed provide a *natural* interferometric WVA of small polarization effect (Faraday rotation). In order to prove this concept, we first develop and experimentally demonstrate a simple yet intuitive interferometric WVA scheme that uses near destructive interference of two paths of an interferometer having slightly rotated linear polarization states of light (playing the role of weak interaction). Using this concept, real and imaginary WVAs of extremely small polarization rotation effect is obtained using small amplitude and phase offsets (mimicking the small post-selection offset parameter $\epsilon$) of the waves in the destructive interference, respectively. The corresponding WVAs are manifested as $\epsilon$-dependent dramatic changes in the polarization state pointer profile, leading to large rotation of the polarization vector orientation angle and changes in the circular polarization descriptor 4$^{th}$ Stokes vector elementof light [27], respectively. We then show that an identical situation of WVA of small Faraday rotation naturally arises near the Fano spectral dip corresponding to the destructive Fano interference between an optically active (having small Faraday rotation) narrow waveguide mode and a polarization isotropic surface plasmon mode in waveguided magneto-plasmonic crystal [24]. Using a theoretical model of interferometric weak value amplification and by employing finite element method (FEM) simulation of Fano resonance in magneto-plasmonic crystal, we demonstrate both real and imaginary WVA of Faraday rotation. This intriguing finding of natural weak value amplification may provide new insights and understanding on a number of non-trivial wave phenomena that originates from fine interference effects.

*Interferometric weak value amplification using polarization state as a pointer*

We consider near destructive interference of two paths of an interferometer having slightly rotated (by a small angle $\alpha$) linear polarizations of light. The corresponding interfering electric field vectors of the two paths are

$$\boldsymbol{E_1} = E_0\hat{\boldsymbol{x}}; \boldsymbol{E_2} = E_0(\cos\alpha\hat{\boldsymbol{x}} + \sin\alpha\hat{\boldsymbol{y}}) \qquad (1)$$

The real and the imaginary WVA of the polarization rotation effect can be obtained by nearly destructive interference of $\boldsymbol{E_1}$ and $\boldsymbol{E_2}$ with small amplitude offset ($\pm\epsilon_a$) and small phase offset ($\pm\epsilon_p$), respectively [2,3]. Here, $\epsilon_a$ is related to the amplitude ratio $a$ of the two paths $\left(\epsilon_a = \frac{1-a}{1+a}\right)$. The corresponding expressions for the resultant electric field for the real and the imaginary WVAs are

$$[(1\pm\epsilon_a)\boldsymbol{E_1} - (1\mp\epsilon_a)\boldsymbol{E_2}]; [e^{\pm i\epsilon_p}\boldsymbol{E_1} - e^{\mp i\epsilon_p}\boldsymbol{E_2}] \qquad (2)$$

The expressions for the $\epsilon$-dependent variations of the resultant electric fields can be worked out using Eq. 1 into Eq. 2. It is convenient to represent the corresponding changes in the polarization state of light (acting as the pointer here) using the intensity-based Stokes vector elements [27]. Using the familiar weak interaction limit ($\alpha \to 0$) it can be shown that the real and the imaginary WVAs of small polarization rotation effect $\alpha$ are manifested as $\epsilon$-dependent changes in the orientation angle ($\psi$) of the polarization vector and the circular (elliptical) polarization descriptor 4$^{th}$ Stokes vector element $\left(\frac{V}{I}\right)$ of light as

$$\psi = \frac{1}{2}\tan^{-1}\left(\frac{U}{Q}\right) \approx \pm \alpha \cot \epsilon_a \quad (3a)$$

$$\frac{V}{I} \approx \pm \alpha \cot \epsilon_p \quad (3b)$$

As evident, as $\epsilon$ becomes small, the polarization vector orientation $\psi$ and the $\frac{V}{I}$ Stokes parameter increase rapidly ($\propto \alpha \cot \epsilon \sim \frac{\alpha}{\epsilon}$ in the limit of small $\epsilon$). Interestingly, like in the conventional linear response regime of WVA [2,4], here also there is a limit on the minimum value of $\epsilon$ ($\epsilon_{min} \sim 2\alpha$) where Eq. 3 is valid, which accordingly sets a bound on the maximum achievable amplification.

We first experimentally demonstrate this WVA concept using a conventional Mach-Zehnder interferometric arrangement (**Figure 1a**). The fundamental Gaussian mode of 632.8 nm line of a He-Ne laser is linearly polarized (horizontal, $x$-polarization) using a polarizer (P1), which then enters the interferometer. A small polarization rotation $\alpha$ (1º) is introduced in one of the arms of the interferometer by orienting the fast axis of a half waveplate (HWP, mounted on a high precision rotational mount) at an angle α/2. The interfering light beams after exiting the interferometer are passed through a combination of a quarter waveplate (QWP) and a linear polarizer (P2) and the resulting interference fringes are imaged into a CCD camera (2048 × 1536 square pixels, pixel dimension 3.45 µm). The quarter waveplate and the linear polarizer combination are used to determine the spatial variation of the four Stokes polarization parameter of light [27]. In order to realize the real WVA scheme, the relative intensities (or amplitude ratio $a$) of light in the two interfering paths were varied using a variable neutral density filter (NDL-25C-2). The polarization vector orientation angle ($\psi$) was determined from the measured $U$ $and$ $Q$ Stokes parameters at a spatial position corresponding to the intensity minima of the interference fringe. The corresponding real WVA was probed by observing changes in the $\psi$-parameter as a function of varying amplitude offset parameter $\epsilon_a$ (or $a$). For probing the imaginary WVA, on the other hand, the intensities of the two paths were kept equal and the $\frac{V}{I}$ Stokes parameter was measured at varying spatial positions away from the point of the intensity minima (which demarcates the position for phase difference $\pi$). The measured spatial variation of $\frac{V}{I}$ was subsequently used to generate its variation as a function of the small phase offset (from $\pi$) parameter $\epsilon_p$.

The experimental results on the interferometric WVA of small polarization rotation effect are summarized in **Figure 1**. The $\frac{V}{I}$ Stokes polarization parameter is observed to exhibit clear increasing trend as one approaches the position of the intensity minima in the fringe pattern (**Fig. 1b**). The variation of $\frac{V}{I}$ with the small phase offset parameter $\epsilon_p$ (**Fig. 1c**) accordingly shows good agreement with the corresponding prediction of Eq. 3b $\left(\propto \alpha \cot \epsilon_p \sim \frac{\alpha}{\epsilon_p}\right)$, demonstrating faithful imaginary WVA. The polarization vector orientation angle ($\psi$), determined from the measured $U$ $and$ $Q$ Stokes parameters also scales as $\left(\propto \alpha \cot \epsilon_a \sim \frac{\alpha}{\epsilon_a}\right)$ with decreasing amplitude offset parameter $\epsilon_a$ (**Fig. 1d**), confirming real WVA of polarization rotation effect $\alpha$ (Eq. 3a). These and other experimental results for varying $\alpha$ validated the concept of interferometric WVA using polarization state as a pointer.

*Natural weak value amplification using spectral interference of Fano resonance*

We now proceed to show that the above interferometric WVA concept *naturally* arises in the spectral interference of optical Fano resonance. For this purpose, we model the electric field of Fano resonance as a coherent superposition of a complex Lorentzian field $(L^R(\omega))$ of a narrow resonance with an ideal continuum [21, 28]. Moreover, we assume that the polarization of the narrow resonance mode is rotated by a small angle $\alpha$ with respect to the continuum mode. The corresponding polarized filed can be expressed as

$$\boldsymbol{E}_s(\omega) \approx [L^R(\omega)(\cos\alpha\,\hat{\boldsymbol{x}} + \sin\alpha\,\hat{\boldsymbol{y}}) + \hat{\boldsymbol{x}}] = \left[\frac{q-i}{\varepsilon(\omega)+i}(\cos\alpha\,\hat{\boldsymbol{x}} + \sin\alpha\,\hat{\boldsymbol{y}}) + \hat{\boldsymbol{x}}\right] \quad (4)$$

Here, $\varepsilon(\omega) = \frac{\omega-\omega_0}{(\gamma/2)}$, $\omega_0$ and $\gamma$ are the central frequency and the width of the narrow resonance and q is the Fano asymmetry parameter. Note that at the Fano frequency $\omega_F =$

$\left(\omega_0 - \frac{q\gamma}{2}\right)$ (corresponding to $\varepsilon = -q$), exact destructive interference takes place, where the phase difference $[\psi(\omega_F)]$ between the narrow resonance and the continuum mode becomes $\pi$ and their amplitude ratio approaches unity ($a = 1$). As one moves *spectrally* slightly away from $\omega_F$ ($\omega = \omega_F \pm \delta$), the desirable near destructive interference scenario is obtained (like in Eq. 2) but now with simultaneous amplitude ($\epsilon_a$) and phase ($\epsilon_p$) offsets. The expressions for these offset parameters can be written as

$$\epsilon_a(\omega) = \frac{1-a}{1+a} = \frac{\left(1-\sqrt{\left(\frac{q^2+1}{\varepsilon^2+1}\right)}\right)}{\left(1+\sqrt{\left(\frac{q^2+1}{\varepsilon^2+1}\right)}\right)}; \quad \epsilon_p(\omega) = \pi - tan^{-1}\left[\frac{q+\varepsilon}{1-q\varepsilon}\right] \quad (5)$$

Using the electric fields of Eq. 4, the expressions for the relevant Stokes vector elements can be derived and the dependence of the resulting $\psi$ and the $\frac{V}{I}$ polarization parameters (Eq. 3) on the $\epsilon_a$ and $\epsilon_p$ parameters can be studied. In **Figure 2**, we have schematically illustrated the spectral analogue of the interferometric WVA concept in Fano resonance (**Fig. 2a**) and presented the corresponding theoretical results using Eq. 4 and Eq. 5 (**Fig. 2b-2d**). As evident, both the polarization vector orientation angle ($\psi$) (**Fig. 2b**) and the $\frac{V}{I}$ Stokes polarization parameter (**Fig. 2c**) exhibit dramatic enhancement at close vicinity of the Fano spectral dip (Energy $E_F = \hbar\omega_F$) corresponding to destructive spectral interference. As one moves *spectrally* away from $E_F$, the $\epsilon_a$ and $\epsilon_p$ parameters change simultaneously and consequently the $\psi$ and the $\frac{V}{I}$ polarization parameters scale as $\left(\sim \alpha cot\, \epsilon_{a/p} \sim \frac{\alpha}{\epsilon_{a/p}}\right)$ simultaneously with both the offset parameters $\epsilon_{a/p}$ (**Fig. 2d**).

We now demonstrate this concept in a realistic Fano resonant system, namely, the waveguided magneto-plasmonic crystals (WMPC) [24-26]. The waveguided plasmonic crystal systems usually comprise of a periodic array of noble metal nanostructures on top of a dielectric waveguiding layer [20-23]. The coupling of the surface plasmons in the metallic nanostructures and the quasi-guided photonic modes in the waveguiding layer leads to Fano resonance [21, 22, 24]. In WMPC, the waveguiding layer is made of magneto optical materials that additionally exhibit Faraday rotation [24, 25], which is the desirable weak interaction parameter in this weak measurement scenario. In what follows, we demonstrate that the recently observed giant enhancement of small Faraday rotation in Fano resonant WMPC [24] is a manifestation of such

*natural* interferometric WVA. For this purpose, we used Finite element method (FEM) [29] for simulating the Fano resonance in the transmitted light from such system (**Figure 3a**). Briefly, the WMPC consists of 1-D periodic grating of gold (Au) metal nanostructures on top of a Y-BIG film, acting as the waveguiding layer [24]. The dimension of the Au grating was ($a$ = 120 nm, $d$ = 495 nm and $h$ = 65 nm) and the thickness of the Y-BIG film was 150 nm. The far field transmission spectra ($\lambda$= 650 nm to 1150 nm) and its polarization dependence were studied. For the simulations, the permittivity tensor elements of the Y-BIG film were taken, $\epsilon_{11} = \epsilon_{33} = 6.7 + 0.053i$ and $g = 0.016 - 0.0092i$ for typical magnetic field of 140mT [24] and dielectric permittivity of Au was taken from literature [30].

**Figure 3** summarizes the corresponding results of natural WVA of Faraday rotation. Significant enhancement of the Faraday rotation (polarization vector orientation angle $\psi$) and the $\frac{V}{I}$ Stokes polarization parameter near the spectral window of the Fano dip are observed (**Fig. 3b**). These variations are subsequently interpreted in terms of the small amplitude and phase offset parameters $\epsilon_a$ and $\epsilon_p$ in the weak value formalism (Eq. 3). In complete agreement with the results of the corresponding theoretical treatment (Fig. 2b-2d), the $\psi$ (**Fig. 3c**) and the $\frac{V}{I}$ Stokes polarization parameter (**Fig. 3d**) simultaneously exhibit *natural* WVA of small Faraday rotation $\left(\sim \alpha \cot \epsilon_{a/p} \sim \frac{\alpha}{\epsilon_{a/p}}\right)$. These results provide conclusive evidence that the the recently observed giant enhancement of small Faraday rotation in Fano resonant WMPC [24] is a manifestation of *natural* interferometric weak value amplification. Note that the simple theoretical model on WVA in Fano resonance additionally predicted that this enhancement of Faraday rotation should be accompanied with evolution of circular polarization as signature of imaginary WVA (Fig. 2c, 2d), which is also verified here by the corresponding results of the FEM based simulation of Fano resonance in WMPC (Fig. 3b and 3d).

In summary, we have developed and experimentally validated a fundamentally interesting concept of interferometric weak value amplification. This concept uses near destructive interference of two paths of an interferometer having slightly rotated linear polarization states of light to obtain both real and imaginary WVA of the small polarization rotation effect. The real and imaginary WVA are manifested as large changes in the polarization state (pointer) of light. Importantly, we have shown that such interferometric WVA naturally arises in Fano resonance due to near destructive spectral interference between a continuum mode and a narrow resonance mode. The corresponding manifestations of natural WVA as giant enhancement of Faraday rotation and dramatic changes in the circular polarization of light are demonstrated in a Fano resonant waveguided magneto-plasmonic crystal system. It is envisaged that with the presence of an appropriate weak interaction parameter such interferometric WVA may naturally arise in diverse other intriguing wave phenomena that originate from fine interference effects. This may open up a new paradigm of natural weak measurements which in turn may provide new insights on rich variety of interference related effects through enhanced probing of fundamentally important weak interactions associated with these effects in the optical domain and beyond.

**Figures**

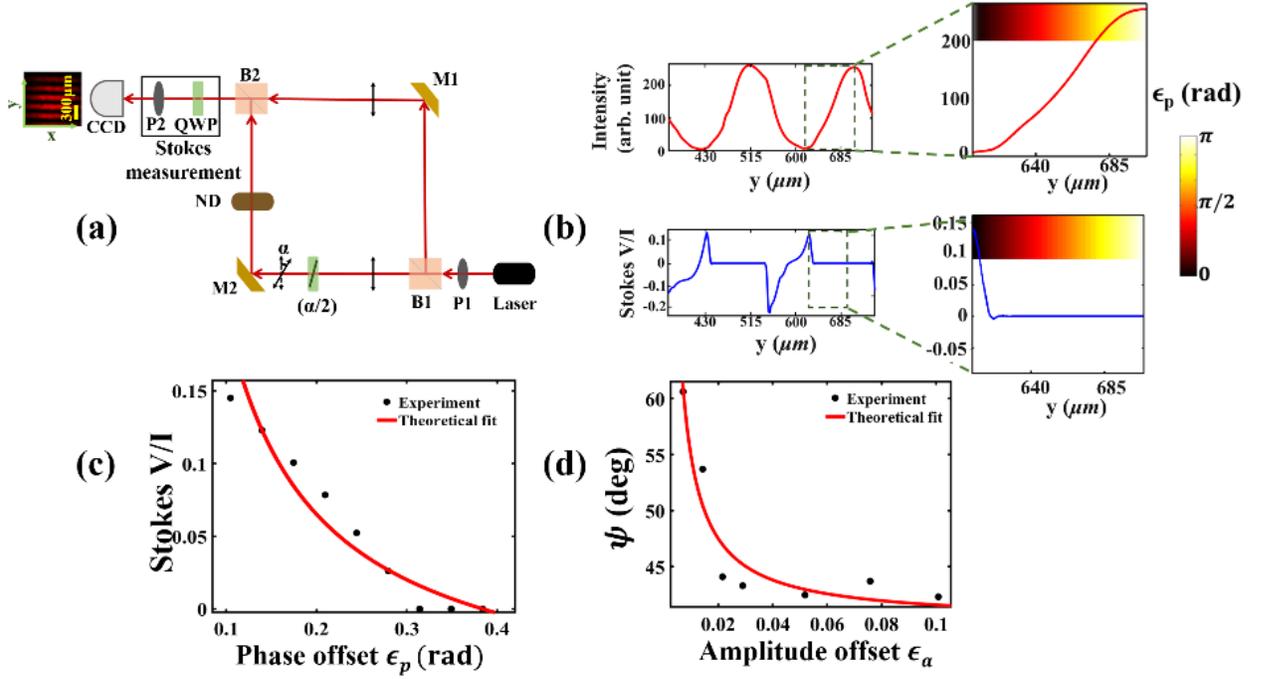

**Figure 1:** *Interferometric WVA using polarization state as a pointer.* (**a**) A schematic of the Mach-Zehnder interferometric arrangement. A half waveplate (HWP) on one of the arms introduces a small polarization rotation $\alpha$ as the weak interaction effect. P1, P2: Glan-Thompson linear polarizers, B1, B2: beam splitters, M1, M2: mirrors, ND: Variable neutral density filter. QWP: Quarter waveplate. (**b**) **and** (**c**): The results of imaginary WVA. (**b**) The spatial (along the *y*-direction shown in the inset of a) variation of the intensity (top panel) and the corresponding $\frac{V}{I}$-Stokes vector element (bottom panel). The corresponding magnified views of the variations around the position of the intensity minima (destructive interference with phase difference $\pi$) are shown along with the phase offset from $\pi$, $\epsilon_p$ (shown by the color bar). (**c**) The corresponding variation of $\frac{V}{I}$ with $\epsilon_p$ (solid circle) and theoretical fit to Eq. 3b (red line). (**d**) The results of real WVA. The variation of the polarization vector orientation angle $\psi$ (solid circle) as a function of the amplitude offset parameter $\epsilon_a$ recorded at the spatial position of intensity minima corresponding to the destructive interference. The corresponding theoretical fit to Eq. 3a is shown by red line.

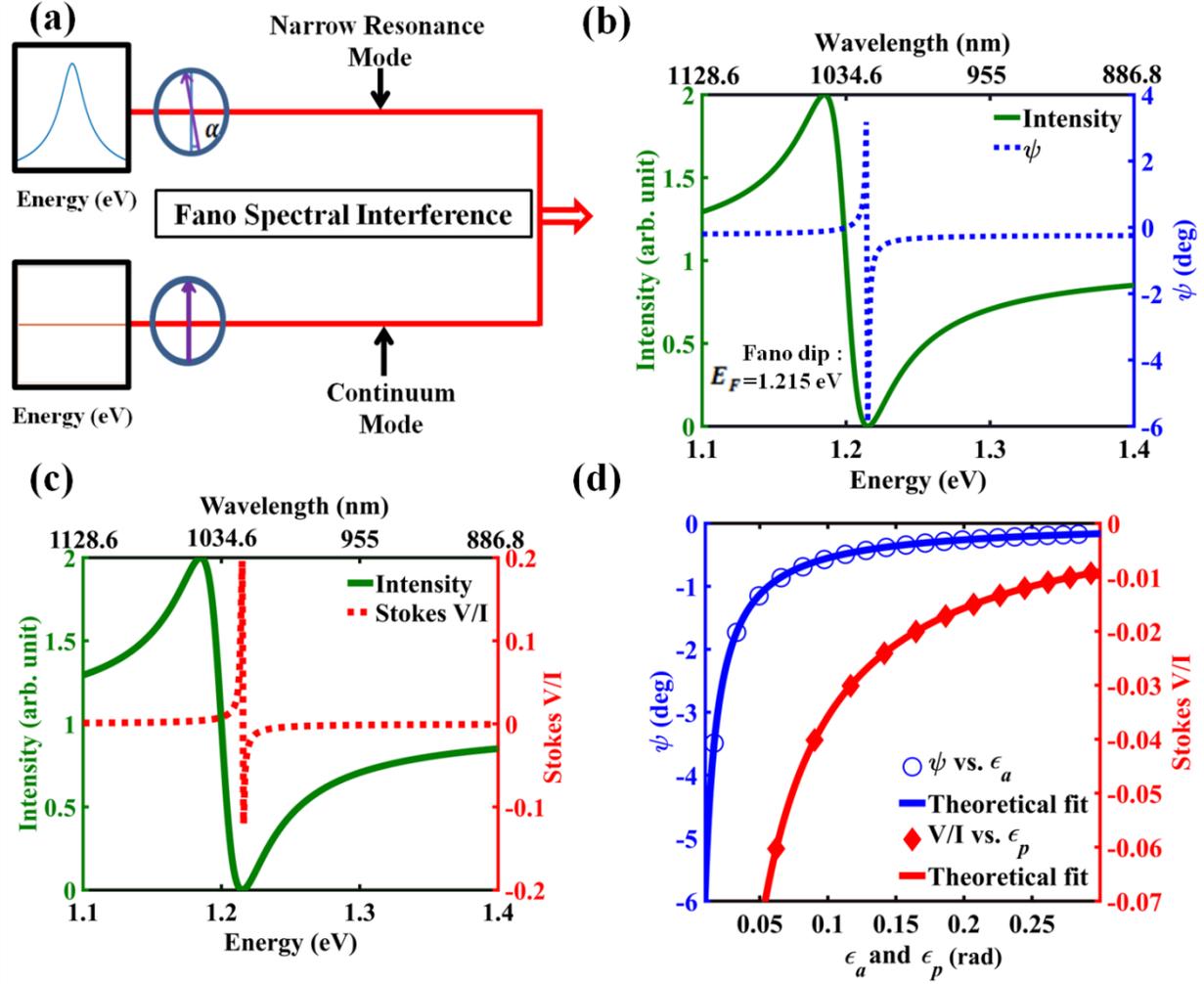

**Figure 2.** *Schematic illustration of natural WVA in the spectral interference of Fano resonance.* (**a**) Interference of a narrow resonance mode with a continuum mode yields an asymmetric Fano spectral line shape of resonance. The small polarization rotation $\alpha$ provides the desirable weak coupling. (**b**) The theoretically calculated (using Eq. 4) spectral variation ($E = \hbar\omega = 1.1 - 1.4$ eV, corresponding to $\lambda = 1128.6 - 886.8$ nm) of the Fano intensity profile (left axis, green solid line) and the polarization vector rotation angle $\psi$ (right axis, blue dashed line). The black arrow denotes the energy (frequency) corresponding to the exact destructive interference or the Fano spectral dip ($E_F = \hbar\omega_F$). (**c**) The variation of the $\frac{V}{I}$ Stokes vector element (right axis, red dashed line). Both the $\psi$ and the $\frac{V}{I}$ polarization parameters exhibit large enhancement as one approaches $E_F$. (**d**) The corresponding variations of the $\psi$ (left axis, blue open circle) and $\frac{V}{I}$ (right axis, red solid diamond) polarization parameters with the amplitude and the phase offset parameters $\epsilon_a$ and $\epsilon_p$ respectively. The theoretical fits to Eq. 3a and 3b are shown by solid lines of the same colors. The following parameters of Fano resonance were used for these theoretical calculations: $E_0 = 1.2\ eV$, $\gamma = 0.03 eV$, $q = -1$, $E_F = 1.215\ eV$ and $\alpha = 0.23$ deg.

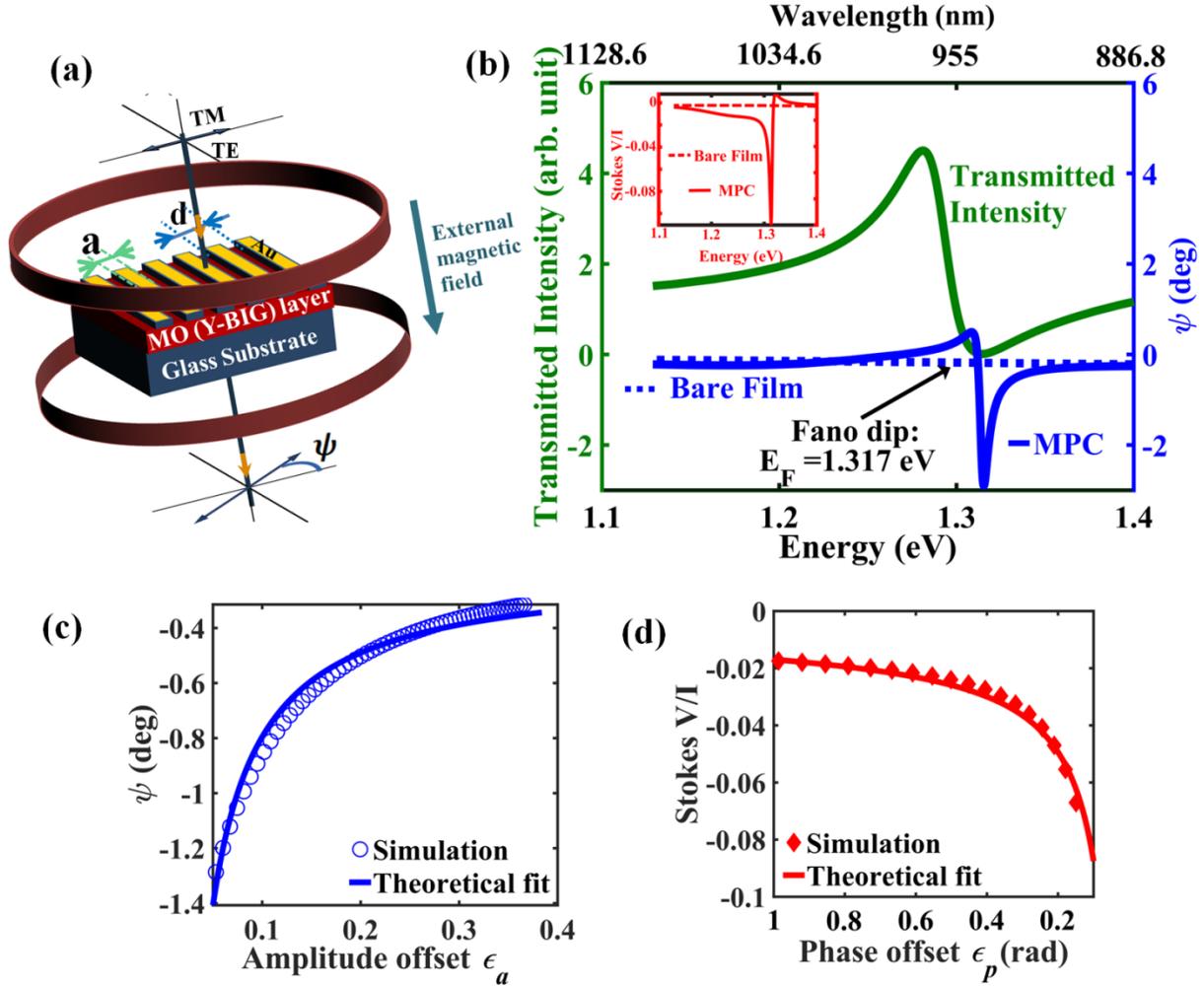

**Figure 3:** *Natural WVA of Faraday rotation in Fano resonant waveguided magneto-plasmonic crystal (WMPC).* **(a)** Schematic illustration of FEM simulation of Fano resonance and Faraday rotation in WMPC. The WMPC system consists of an Au grating on top of a thin (Y-BIG) film. The grating parameters are *a*- width, d- centre to centre distance, *h*- height. **(b)** The transmitted intensity (left axis, green line) exhibits prominent Fano spectral asymmetry ($E = \hbar\omega = 1.1 - 1.4$ eV, corresponding to $\lambda = 1128.6 - 886.8$ nm shown here). The spectral variation of the polarization vector rotation angle $\psi$ (enhanced Faraday rotation) of the WMPC (right axis, blue solid line) is shown along with the corresponding rotation for a bare Y-BIG film (right axis, blue dashed line). The corresponding variation of the $\frac{V}{I}$ Stokes polarization parameter is shown in the inset. **(c)** The natural WVA of the polarization vector rotation angle $\psi$ (Faraday rotation) (blue open circles) as a function of amplitude offset parameter $\epsilon_a$. **(d)** The corresponding natural WVA of the $\frac{V}{I}$ parameter (red solid diamond) as a function of phase offset parameter $\epsilon_p$. The theoretical fits to Eq. 3a and 3b are shown by solid lines of same colors. The chosen morphological parameters of the WMPC (see text) resulted in the following parameters of Fano resonance: $E_0 = 1.291 \ eV$, $\gamma = 0.0323 \ eV$, $q = -1.612$, $E_F = 1.317 \ eV$.